# On the local discrimination of quantum measurements

Youbang Zhan

*School of Physics and Electronic Electrical Engineering,*
*Huaiyin Normal University, Huaian 223300, P. R. China*

E-mail address:  youbangzhan@163.com

ABSTRACT

  The discrimination of quantum measurements is an important subject of quantum information processes. In this paper we present a novel protocol for local quantum measurement discrimination (LQMD) with multi-qubit entanglement systems. It is shown that, in the case of Bob did not know Alice's result of measurement, if both two observers (Alice and Bob) agreed in advance that Alice should measure her qubits before an appointed time t, the local discrimination of two different kinds of measurement can be completed via numerous eight-qubit GHZ entangled states and selective projective measurement.

**1. Introduction**

   Quantum entanglement is one of the striking features of quantum mechanics [1]. The nonlocal nature of entanglement is the essential resource for many quantum information tasks including teleportation [2] and super-dense coding [3]. However, although entanglement appears to allow particles which are separated in space to influence one another instantaneously [4], it has been pointed that this cannot be used to signal without help of classical communication [5-9].
    On the other hand, it is well-known that measurement is a central tenet of quantum mechanics. The problem of discrimination between quantum measurements has been recently considered in quantum information tasks [10-13]. Ji *et al*. [10] have proposed simple schemes that can perfectly identify projective measurement apparatuses secretly chosen from a finite set. Entanglement is used in this scheme both to make possible the perfect identification and to improve the efficiency significantly. Fiurasek and Micuda [11] have studied optimal discrimination between two projective quantum measurements on a single qubit. Ziman *et al* . [12] have investigated the unambiguous comparison of unknown quantum measurements represented by non-degenerate sharp positive operator valued measures (POVM). One can notice that, in above works [10-13] of discriminating quantum measurement, employing classical communication is necessary.
   In the last decade, the impossibility of the local quantum measurement discrimination (LQMD) has been frequently discussed and proved (*e. g.* [14-19]). For example, assume that Alice and Bob share bipartite quantum system described by a known state $\rho$ [19]. They can make local measurements, with elements

$$\sum_i A_i^\dagger A_i = I, \quad \sum_j B_j^\dagger B_j = I \tag{1}$$

on the subsystems $A$ and $B$ respectively, where $A_i$ and $B_j$ are the "detector operators" associated to the elements of a POVM for the observation of results $\mu_A$ by Alice and $\nu_B$ by Bob. If Bob is not informed that Alice got outcome $\mu_A$, the mean value that he gets any observable $\nu_B$ is

$$\langle \nu_B \rangle = tr_{AB}\left\{\sum_i A_i \rho A_i^\dagger \nu_B\right\} = tr_A\left\{\sum_i A_i^\dagger A_i tr_B\{\rho \nu_B\}\right\}$$
$$= tr_B\{tr_A\{\rho\} \nu_B\} . \tag{2}$$

Since the result of Eq. (2) does not depend on Alice's operators, Bob cannot decide what measurements Alice did, or worse, he cannot even tell if she had measured or not, without her help [19]. Obviously, in this example, Alice not only did not inform Bob of her measurement result, and even did not declare that whether she had measured.

Different from above example [19], here we will discuss another case for LQMD. In this case, Bob could know that Alice had completed the measurement after her operation and did not know Alice's result of measurement. By a careful analysis, we find that if multiple multi-qubit entangled states and a kind of special measurement (called selective projective measurement (SPM)) are employed, the local discrimination of quantum measurements can be realized without assistance of classical communication. In this work, we present a novel protocol for LQMD via selective projective measurement with numerous eight-qubit GHZ states. It is shown that, in the case of Bob did not know Alice's result of measurement, if both two observers (Alice and Bob) agreed in advance that Alice should measure her qubits by using the SPMs before an appointed time (it is equivalent that, after her measurement, Alice only announced publicly that she had completed the measurement, and did not declare the result of her measurement), the local discrimination of two different kinds of measurement can be realized by using a series of single-qubit correlative measuring basis.

## 2. Two different kinds of quantum projective measurement

Suppose that an eight-qubit GHZ state is shared by Alice and Bob,

$$|\Phi\rangle = \frac{1}{\sqrt{2}}(|00000000\rangle + |11111111\rangle)_{A_1 A_2 A_3 A_4 A_5 A_6 A_7 B} , \tag{3}$$

where qubits $A_1$, $A_2$, ..., $A_7$ are in the possession of Alice and $B$ belongs to Bob. Assume that Alice and Bob agreed in advance that Alice should measure her qubits before an appointed time. Now, let Alice make two different kinds of measurement on the state $|\Phi\rangle$. In the first kind of measurement, Alice makes common projective measurements (CPMs) on her qubits $A_1$, $A_2$, ...,and $A_7$ under the

measurement basis $\{|+\rangle, |-\rangle\}$, where $|+\rangle = \frac{1}{\sqrt{2}}(|0\rangle + |1\rangle)$, $|-\rangle = \frac{1}{\sqrt{2}}(|0\rangle - |1\rangle)$, successively.

One can see that, after measurements of Alice, 128 possible final collapsed states of the qubit $B$ will always be $\frac{1}{8\sqrt{2}}|+\rangle_B$ or $\frac{1}{8\sqrt{2}}|-\rangle_B$. Now we turn to the second kind of measurement. To realize the LQMD, Alice will utilize a novel kind of projective measurements, which we refer to as SPMs, with a series of single-qubit correlative measuring basis, on her qubits. Firstly, Alice measures the qubit $A_1$ in the state $|\Phi\rangle$ under the basis $\{|v\rangle, |v^\perp\rangle\}$, where $|v\rangle = x|0\rangle + y|1\rangle$, $|v^\perp\rangle = y|0\rangle - x|1\rangle$, $x$ and $y$ are real, $x^2 + y^2 = 1$, and let $x = \sqrt{6}/3$, $y = \sqrt{3}/3$. If measurement outcome of Alice is $|v\rangle_{A_1}$, the state of qubits $A_2$, $A_3$, ..., $A_7$ and $B$ will evolve as

$$|\phi_1\rangle = \frac{1}{\sqrt{2}F_1}\left(x|0000000\rangle + y|1111111\rangle\right)_{A_2 A_3 A_4 A_5 A_6 A_7 B}, \qquad (4)$$

where we let $F_1 = 1$, Alice can in turn measure the qubits $A_2$, $A_3$, ..., $A_7$ under the basis $\{|+\rangle, |-\rangle\}$. After that, the qubit $B$ will always be in the state $\frac{1}{8}|\mu^+\rangle_B$ or $\frac{1}{8}|\mu^-\rangle_B$, here $|\mu^+\rangle = \frac{1}{\sqrt{2}}(x|0\rangle + y|1\rangle)$ and $|\mu^-\rangle = \frac{1}{\sqrt{2}}(x|0\rangle - y|1\rangle)$. If measurement result of Alice is $|v^\perp\rangle_{A_1}$, the qubits $A_2$, $A_3$, ..., $A_6$ and $B$ will be in the state of

$$|\phi_1'\rangle = \frac{1}{\sqrt{2}F_1}\left(y|0000000\rangle - x|1111111\rangle\right)_{A_2 A_3 A_4 A_5 A_6 A_7 B}. \qquad (5)$$

Then Alice measures the qubit $A_2$ under the measurement basis $\{|\lambda_1\rangle, |\lambda_1^\perp\rangle\}$, which is given by

$$|\lambda_1\rangle = \frac{1}{F_2}\left(\frac{x}{y}|0\rangle + \frac{y}{x}|1\rangle\right), \qquad |\lambda_1^\perp\rangle = \frac{1}{F_2}\left(\frac{y}{x}|0\rangle - \frac{x}{y}|1\rangle\right), \qquad (6)$$

where $F_2 = \left[(x/y)^2 + (y/x)^2\right]^{1/2}$. Corresponding to Alice's measurement outcome $|\lambda_1\rangle_{A_2}$ or $|\lambda_1^\perp\rangle_{A_2}$, the state of qubits $A_3$, ..., $A_7$ and $B$ will evolve as $|\phi_2\rangle$ or $|\phi_2'\rangle$, which can be expressed as

$$|\phi_2\rangle = \frac{1}{\sqrt{2}F_2}\left(x|000000\rangle - y|111111\rangle\right)_{A_3 A_4 A_5 A_6 A_7 B},$$

$$|\phi_2'\rangle = \frac{1}{\sqrt{2}F_2}\left(\frac{y^2}{x}|000000\rangle + \frac{x^2}{y}|111111\rangle\right)_{A_3A_4A_5A_6A_7B} \tag{7}$$

As described above, we can easy find that the goal of the SPMs is as much as possible to make the qubit $B$ collapsed into the state $\frac{1}{R}|\mu^+\rangle$ or $\frac{1}{R}|\mu^-\rangle$ after all, where $R$ is a constant or a coefficient related to $x$ and $y$. By the formulae deducing, a detailed implementation procedure for the SPM has been provided and 128 possible final collapsed states of the qubit $B$ after Alice's measurements are given in Appendix A. The relation of the results of Alice's measurement and 128 possible final collapsed states of the qubit $B$ can be expressed as follows:

$$|v\rangle_{A_1} \rightarrow |\psi_1^\pm\rangle_B = \frac{1}{8T_1}|\mu^\pm\rangle_B \quad \text{(64 terms)}$$

$$|\lambda_1\rangle_{A_2} \rightarrow |\psi_2^\pm\rangle_B = \frac{1}{4\sqrt{2}T_2}|\mu^\pm\rangle_B \quad \text{(32 terms)}$$

$$|\lambda_2\rangle_{A_3} \rightarrow |\psi_3^\pm\rangle_B = \frac{1}{4T_3}|\mu^\pm\rangle_B \quad \text{(16 terms)}$$

$$|\lambda_3\rangle_{A_4} \rightarrow |\psi_4^\pm\rangle_B = \frac{1}{2\sqrt{2}T_4}|\mu^\pm\rangle_B \quad \text{(8 terms)}$$

$$|\lambda_4\rangle_{A_5} \rightarrow |\psi_5^\pm\rangle_B = \frac{1}{2T_5}|\mu^\pm\rangle_B \quad \text{(4 terms)}$$

$$|\lambda_5\rangle_{A_6} \rightarrow |\psi_6^\pm\rangle_B = \frac{1}{\sqrt{2}T_6}|\mu^\pm\rangle_B \quad \text{(2 terms)}$$

$$|\lambda_6\rangle_{A_7} \rightarrow |\psi_7^+\rangle_B = \frac{1}{T_7}|\mu^+\rangle_B \quad \text{(1 term)}$$

$$|\lambda_6^\perp\rangle_{A_7} \rightarrow |\psi_7^-\rangle_B = P|\eta\rangle_B, \quad \text{(1 term)} \tag{8}$$

where $T_m = F_1F_2\cdots F_m \ (m=1,2,\cdots,7)$, and

$$F_1 = 1,$$

$$F_2 = \left[(x/y)^2 + (y/x)^2\right]^{1/2},$$

$$F_3 = \left[(x/y)^4 + (y/x)^4\right]^{1/2},$$

$$F_4 = \left[(x/y)^8 + (y/x)^8\right]^{1/2},$$

$$F_5 = \left[ (x/y)^{16} + (y/x)^{16} \right]^{1/2} ,$$

$$F_6 = \left[ (x/y)^{32} + (y/x)^{32} \right]^{1/2} ,$$

$$F_7 = \left[ (x/y)^{64} + (y/x)^{64} \right]^{1/2} , \tag{9}$$

and $P = \dfrac{\sqrt{x^{254} + y^{254}}}{\sqrt{2} T_7 x^{63} y^{63}}$, $|\eta\rangle_B$ is a normalized state, which is given by

$$|\eta\rangle_B = \frac{1}{\sqrt{x^{254} + y^{254}}} \left( y^{127} |0\rangle - x^{127} |1\rangle \right)_B . \tag{10}$$

Thus much Alice's selective measurements have been completed. From Eqs. (8) - (10), it is easy noted that, after Alice performing the SPMs on her all qubits, the states $\dfrac{1}{g_n T_n} |\mu^{\pm}\rangle_B$ ( $g_n = 2^{(7-n)/2}$, $n = 1, 2, \cdots, 7$ ) in all 128 collapsed states of the qubit $B$ accounted for 127, and the state $|\psi_7^-\rangle_B$ for 1. On the other hand, by simple calculation, one can find that, after Alice's measurements the probability of the qubit $B$ being in the state $\dfrac{1}{g_n T_n} |\mu^{\pm}\rangle_B$ ( $g_n = 2^{(7-n)/2}$, $n = 1, 2, \cdots, 7$ ) is 0.75, and in the state $|\eta\rangle_B$ is 0.25. It must be pointed out that it is just these measured results of the SPM that led to the realization of the LQMD. Figure 1 shows the detailed configuration of the CPM and SPM.

Clearly, after Alice performing the CPMs or SPMs on her qubits respectively, the final collapsed states of the qubit $B$ are obvious different. As mentioned above, if Alice makes the CPMs on her qubits, after her measurements, 128 possible final collapsed states of the qubit $B$ will always be $\dfrac{1}{8\sqrt{2}} |+\rangle_B$ or $\dfrac{1}{8\sqrt{2}} |-\rangle_B$. If Alice employs the SPMs on her qubits, after her measurements, 128 possible final collapsed states of the qubit $B$ can be given by Eq. (8). It must be emphasized that, whether Alice's measurements are the CPMs or SPMs, since Alice and Bob agreed in advance that Alice should measure her qubits before an appointed time, Bob can always know that the qubit $B$ must be collapsed into the state corresponded to one of Alice's 128 results of measurement after Alice's measurements.

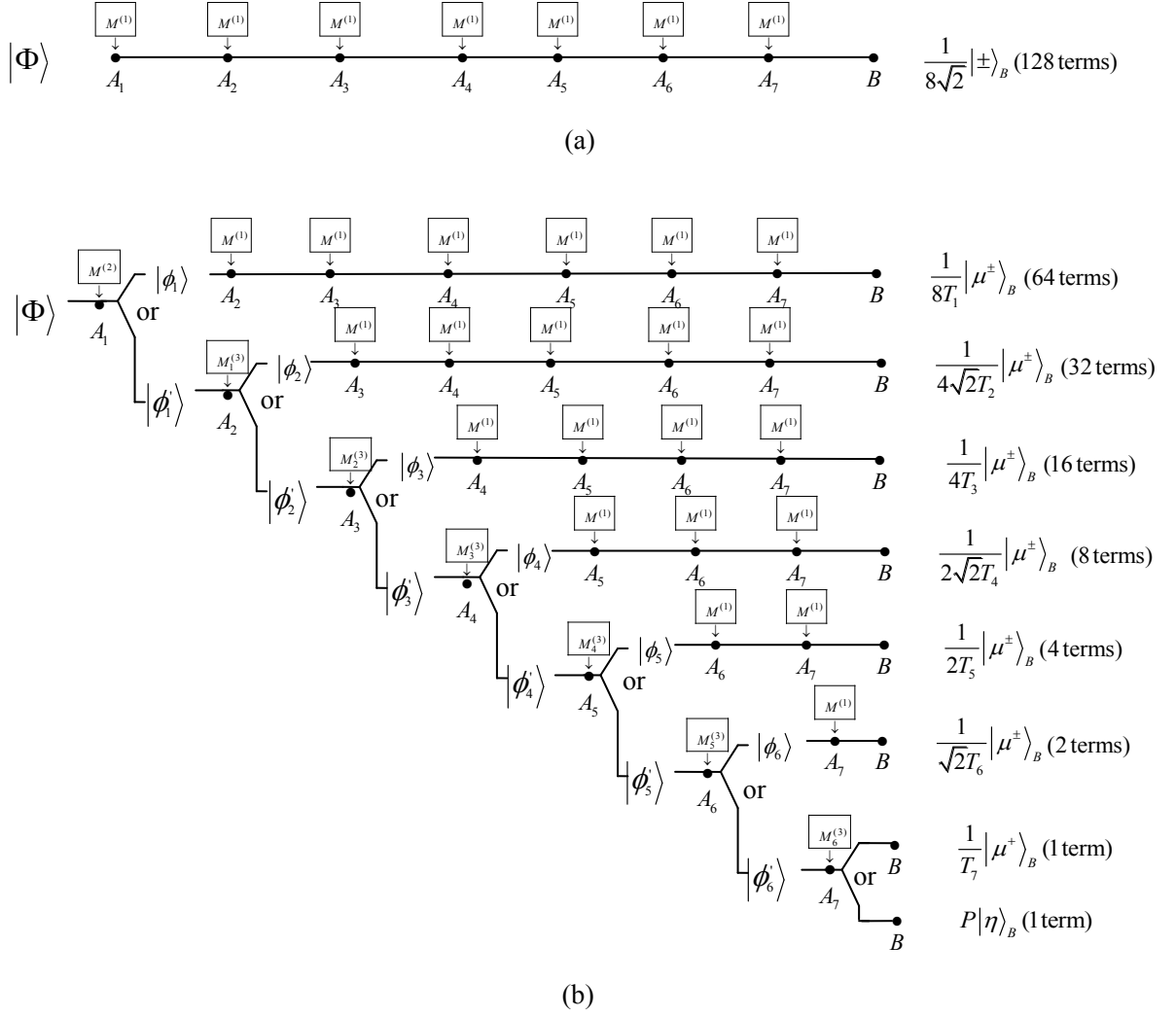

Fig.1 A sketch of the CMP (a) and SPM (b). Each dot denotes a qubit. $M^{(1)}(M^{(2)}, M_i^{(3)}(i=1,2,\cdots$ denotes Alice's single-qubit projective measurement on qubits $A_1, A_2, \cdots$ under measuring basis $\{|+\rangle, |-\rangle\}(\{|\nu\rangle, |\nu^\perp\rangle\}, \{|\lambda_i\rangle, |\lambda_i^\perp\rangle\})(i=1,2,\cdots$

## 3. Local discrimination of two different kinds of measurement with numerous eight-qubit GHZ states

The detailed procedure of our LQMD protocol can be described as follows. Suppose that two space-like separated observers, Alice and Bob, share $N$ eight-qubit GHZ states. To ensure the following analysis becomes exact, here we take $N=30$ [20]. Thus, the 30 eight-qubit GHZ states can be given by

$$|\Phi^{(k)}\rangle = \frac{1}{\sqrt{2}}(|00000000\rangle + |11111111\rangle)_{A_1^{(k)} A_2^{(k)} A_3^{(k)} A_4^{(k)} A_5^{(k)} A_6^{(k)} A_7^{(k)} B^{(k)}}, \qquad (11)$$

where $k=1,2,\cdots,30$, and the qubits $A_1^{(k)}$, $A_2^{(k)}$, ..., $A_7^{(k)}$ are in the possession of Alice and $B^{(k)}$ belong to Bob. Different from previous quantum operation discrimination schemes, we assume that there is no classical channel between Alice and Bob, and Bob is not informed that Alice got her result of measurement. In this case, before the agreed time $t$, Alice should randomly make two different kinds of measurements, CPMs or SPMs, on her qubits in the state $|\Phi^{(k)}\rangle$ ($k=1,2,\cdots,30$) respectively. Now we will consider the local discrimination of two different measurements.

(s1) If Alice performs the CPMs on her qubits, after Alice's measurements, all qubits $B^{(k)}$ will be in the states $\frac{1}{8\sqrt{2}}|+\rangle_{B^{(k)}}$ or $\frac{1}{8\sqrt{2}}|-\rangle_{B^{(k)}}$. At the appointed time $t$, Bob measures his qubits $B^{(k)}$ all in the basis $\{|0\rangle,|1\rangle\}$. After Bob's measurements, by statistics theory, the probability of all qubits $B^{(k)}$ in the state $|0\rangle$ or $|1\rangle$ will be in the ratio of one to one.

(s2) If Alice's measurements are the SPMs, by mentioned above, after Alice's selective measurements, the probability of all qubits $B^{(k)}$ in the states $\frac{1}{g_n T_n}|\mu^+\rangle_B$ or $\frac{1}{g_n T_n}|\mu^-\rangle_B$ ($g_n = 2^{(7-n)/2}$, $n=1,2,\cdots,7$) is $(0.75)^{30} \approx 0.00018$, *i.e.*, the probability of at least one qubit $B^{(k')}$ in the state $|\psi_7^-\rangle_B$ is $1-(0.75)^{30} \approx 0.99982$. This means that, after Alice's SPMs, at least one qubit $B^{(k)}$ will be collapsed into the state $|\psi_7^-\rangle_B$. Then, at the appointed time $t$, Bob measures the qubits $B^{(k)}$ all in the basis $\{|0\rangle,|1\rangle\}$. One can find that, after Bob's measurements, the probability of the qubits $B^{(k)}$ in the state $|0\rangle$ or $|1\rangle$ will be different from the case Alice employed the CPMs. To illustrate this clearly, without loss of generality, we first discuss the situation in which only one qubit $B^{(k')}$ in the state $|\psi_7^-\rangle_B$ after Alice's measurements. From the state $|\psi_7^-\rangle_B$ in Eq. (8), it is easily found that, after measurements of Bob, the probability of the qubit $B^{(k')}$ in the state $|0\rangle$ or $|1\rangle$ will be in the ratio of one to $u$ ($u = \left(\alpha^{127}/\beta^{127}\right)^2 \approx 1.7\times 10^{38}$), that is, the qubit $B^{(k')}$ will be always collapsed into the state $|1\rangle$. As a special case, we also assume that all the other 29 qubits $B^{(k)}$ are in the states $|\psi_1^\pm\rangle_B$ after Alice's measurements and then all the 29 qubits are in the state $|0\rangle$ after Bob's measurements. In this case, by simple calculation, one can easily find that the probability of the 30 qubits $B^{(k)}$ in the state $|0\rangle$ or $|1\rangle$ will be in the ratio of 1 to $w_{(1)}$ after Bob's measurements, here $w_{(1)}$ is given by

$$w_{(1)} = \left(\frac{x^{127}}{\sqrt{2}T_7 x^{63} y^{63}}\right)^2 \bigg/ \left[29\left(\frac{x}{8T_1\sqrt{2}}\right)^2\right] \approx 1.655 \tag{12}$$

For general cases in which only one qubit $B^{(k')}$ in the state $|\psi_7^-\rangle_B$ and other 29 qubits $B^{(k)}$ collapsed randomly into the states $\frac{1}{g_n T_n}|\mu^\pm\rangle_B$ ( $g_n = 2^{(7-n)/2}$, $n = 1,2,...,7$) after Alice's measurements, it is easily found that the probability of the 30 qubits $B^{(k)}$ in the state $|0\rangle$ or $|1\rangle$ will be in the ratio of one to $w_{(1)}$ ($w_{(1)} > 1.655$) after Bob's measurements.

(s3) Now let us discuss the situation in which there are two qubits $B^{(k')}$ and $B^{(k'')}$ in the state $|\psi_7^-\rangle$ after Alice's measurements. Similar to the above mentioned, one can see that the probability of the 30 qubits $B^{(k)}$ in the state $|0\rangle$ or $|1\rangle$ will be in the ratio of one to $w_{(2)}$ ($w_{(2)} \geq 3.43$) after Bob's measurements.

(s4) For the cases in which more qubits $B^{(1)}$, $B^{(2)}$, ..., $B^{(l)}$ ($l = 3,4,\cdots,30$) collapsed into the state $|\psi_7^-\rangle_B$ after Alice's measurements, the probability of the 30 qubits $B^{(k)}$ in the state $|0\rangle$ or $|1\rangle$ will be in the ratio of one to $w_{(l)}$ ($w_{(l)} > w_{(2)}$, $l = 3,4,\cdots,30$) after Bob's measurements.

As described above, after measurements of Alice, the probability of the 30 qubits $B^{(k)}$ in the state $|0\rangle$ or $|1\rangle$ will be in the ratio of one to $W$ ($W \geq 1.655$) (we call $W$ the discriminated parameter) after Bob's measurements, where $W \in \{w_{(j)} : j = 1,2,\cdots,30\}$.

(s5) To ensure the outcome of Bob's measurements more reliable, it can be further supposed that Alice and Bob share 20 entangled states groups (ESGs), each consisting of 30 eight-qubit GHZ states $|\Phi^{(k)}\rangle$ (see Eq. (11)). If Alice's measurements are the CPMs, it is easy found that, after Alice's and Bob's measurements, the probability of all qubits $B^{(k)}$ of each ESG in the state $|0\rangle$ or $|1\rangle$ will be still in the ratio of one to one. If Alice's measurements are the SPMs, by statistics theory, after Alice's and Bob's measurements, in all ESGs the probability of the qubits $B^{(k)}$ of each ESG in the state $|0\rangle$ or $|1\rangle$ will be in the ratio of one to $W$ ($W \geq 1.655$).

As mentioned above, one can see that, in this protocol, at the appointed time $t$, Bob should measure his qubits $B^{(k)}$ all in the basis $\{|0\rangle, |1\rangle\}$. If Alice performs the CPMs on her qubits, after Bob's measurements, the probability of all qubits $B^{(k)}$ in the state $|0\rangle$ or $|1\rangle$ will be in the ratio of one to one. If Alice's measurements are the SPMs, after Bob's measurements, the probability of the qubits $B^{(k)}$ of each ESG in the state $|0\rangle$ or $|1\rangle$ will be in the ratio of one to $W$ ($W \geq 1.655$). According to these results, Bob can distinguish that the measurements used by Alice are CPMs or SPMs. Thus, the LQMD is realized successfully.

**4. Discussion and conclusion**

Before conclusion, we make some discussion. (i) It should be noted that, in the present LQMD protocol, Bob did not obtain Alice's quantum information, *i.e.*, if Alice's measurements are SPMs, Bob couldn't have learned the coefficients $x$ and $y$ in the measuring basis performed by Alice since he is not informed that Alice got result of measurement. In fact, Bob doesn't need to know Alice's quantum information (*e.g.* the coefficients $x$ and $y$). As mentioned above, after his measurements, Bob can determine that the measurements performed by Alice are CPMs or SPMs only according to the probability of his qubits $B^{(k)}$ in the state $|0\rangle$ or $|1\rangle$. That is to say, in our LQMD protocol, the entanglement can be used for transmission of information (*e.g.* the classical messages 0 and 1 can be represented by CPMs and SPMs respectively) without assistance of classical communication. (ii) It must be pointed that, in our protocol, it is essential that eight-particle GHZ states are applied. It is easy found that if $l$-particle GHZ states ($l < 8$) are employed, the LQMD will not be completed. For example, if 30 seven- or six-particle GHZ states are used, from (s2) in section 4 one can see that, the discriminated parameter $W$ will be 0.83 or 0.41. In this case, the CPMs and SPMs cannot be distinguished. On the other hand, to ensure the discriminated parameter $W \geq 1.655$, one can only use 15 seven-particle or 7 (7.5) six-particle GHZ states. However, in these cases, the exact of measurement results will not be guaranteed. It is just because of that numerous eight-qubit GHZ states and the SPMs have been used, our LQMD protocol can be completed successfully. (iii) We should emphasize that our work has been completed in the framework of standard quantum mechanics.

In conclusion, we have proposed a theoretical protocol for local discrimination of two different kinds of measurement by using selective measurement and numerous eight-qubit GHZ states. To realize the protocol, a series of single-qubit correlative measuring basis has been employed. It is shown that, in the case of Bob is not informed that Alice got her result of measurement, if both two observers agreed in advance that Alice should measure her qubits before an appointed time, the LQMD can be realized successfully without assistance of classical information. So far there has been experiment implementing the eight-qubit GHZ state [21], hence, we hope our work can be experimentally realized in the near future and stimulate further research on quantum communication and quantum information processing.

**Acknowledgements**

The author wishes to thank Hou-Fang Mu, Yong-Sheng Zhang, Yan-Xiao Gong, Chuan-Zhi Bai, Peng-Cheng Ma and Qun-Yong Zhang for useful discussions and assistances.**Appendix A**

From Eqs. (6) and (7), Alice can measure her qubits according to the result of her own measurement. If result of Alice's measurement is $|\lambda_1\rangle_{A_2}$ in state (6), she should measure her qubits $A_3, \ldots, A_7$ in state $|\phi_2\rangle$ (see Eq. (7)) under the basis $\{|+\rangle,|-\rangle\}$, successively. After that, the qubit $B$ will always be in the state

$$|\psi_2^+\rangle_B = \frac{1}{4\sqrt{2}T_2}|\mu^+\rangle_B \text{ or } |\psi_2^-\rangle_B = \frac{1}{4\sqrt{2}T_2}|\mu^-\rangle_B. \tag{A1}$$

If Alice's measured outcome is $|\lambda_1^\perp\rangle_{A_2}$, she can measure her qubit $A_3$ in state $|\phi_2'\rangle$ (see Eq. (7)) under the basis $\{|\lambda_2\rangle,|\lambda_2^\perp\rangle\}$, which is given by

$$|\lambda_2\rangle = \frac{1}{F_3}\left(\frac{x^2}{y^2}|0\rangle + \frac{y^2}{x^2}|1\rangle\right),$$

$$|\lambda_2^\perp\rangle = \frac{1}{F_3}\left(\frac{y^2}{x^2}|0\rangle - \frac{x^2}{y^2}|1\rangle\right), \tag{A2}$$

where $F_3 = \left[(x/y)^4 + (y/x)^4\right]^{1/2}$. If Alice's result of measurement is $|\lambda_2\rangle_{A_3}$, the qubits $A_4, \ldots, A_7$ and $B$ will be collapsed into the state $|\phi_3\rangle$, which is given by

$$|\phi_3\rangle = \frac{1}{\sqrt{2}T_3}\left(x|00000\rangle + y|11111\rangle\right)_{A_4A_5A_6A_7B}, \tag{A3}$$

where $T_3 = F_1F_2F_3$. Then Alice can in turn measure her qubits $A_4, \ldots, A_7$ in the basis $\{|+\rangle,|-\rangle\}$, and qubit $B$ will be collapsed into the state

$$|\psi_3^+\rangle_B = \frac{1}{4T_3}|\mu^+\rangle_B \text{ or } |\psi_3^-\rangle_B = \frac{1}{4T_3}|\mu^-\rangle_B. \tag{A4}$$

If Alice's outcome of measurement is $|\lambda_2^\perp\rangle_{A_3}$, the state of qubits $A_4, \ldots, A_7$ and $B$ will evolve as

$$|\phi_3'\rangle = \frac{1}{\sqrt{2}T_3}\left(\frac{y^4}{x^3}|00000\rangle - \frac{x^4}{y^3}|11111\rangle\right)_{A_4A_5A_6A_7B}. \tag{A5}$$

Then Alice can measure her qubit $A_4$ in the basis

$$|\lambda_3\rangle = \frac{1}{F_4}\left(\frac{x^4}{y^4}|0\rangle + \frac{y^4}{x^4}|1\rangle\right),$$

$$|\lambda_3^\perp\rangle = \frac{1}{F_4}\left(\frac{y^4}{x^4}|0\rangle - \frac{x^4}{y^4}|1\rangle\right), \quad (A6)$$

where $F_4 = \left[(x/y)^8 + (y/x)^8\right]^{1/2}$. If Alice's result of measurement is $|\lambda_3\rangle_{A_4}$, the qubits $A_5$, $A_6$, $A_7$ and $B$ will be in the state of

$$|\phi_4\rangle = \frac{1}{\sqrt{2}T_4}\left(x|0000\rangle - y|1111\rangle\right)_{A_5 A_6 A_7 B}, \quad (A7)$$

where $T_4 = F_1 F_2 F_3 F_4$. Alice should measure her qubits $A_5$, $A_6$ and $A_7$ in the basis $\{|+\rangle, |-\rangle\}$, then qubit $B$ will be in the state

$$|\psi_4^+\rangle_B = \frac{1}{2\sqrt{2}T_4}|\mu^+\rangle_B \quad \text{or} \quad |\psi_4^-\rangle_B = \frac{1}{2\sqrt{2}T_4}|\mu^-\rangle_B. \quad (A8)$$

If Alice's outcome of measurement is $|\lambda_3^\perp\rangle_{A_4}$, the state of qubits $A_5$, $A_6$, $A_7$ and $B$ will be transferred as

$$|\phi_4'\rangle = \frac{1}{\sqrt{2}T_4}\left(\frac{y^8}{x^7}|0000\rangle + \frac{x^8}{y^7}|1111\rangle\right)_{A_5 A_6 A_7 B}. \quad (A9)$$

Alice can measure her qubit $A_5$ under the basis

$$|\lambda_4\rangle = \frac{1}{F_5}\left(\frac{x^8}{y^8}|0\rangle + \frac{y^8}{x^8}|1\rangle\right),$$

$$|\lambda_4^\perp\rangle = \frac{1}{F_5}\left(\frac{y^8}{x^8}|0\rangle - \frac{x^8}{y^8}|1\rangle\right), \quad (A10)$$

where $F_5 = \left[(x/y)^{16} + (y/x)^{16}\right]^{1/2}$. If Alice's result of measurement is $|\lambda_4\rangle_{A_5}$, the state of qubits $A_6$, $A_7$ and $B$ will evolve as

$$|\phi_5\rangle = \frac{1}{\sqrt{2}T_5}\left(x|000\rangle + y|111\rangle\right)_{A_6 A_7 B}, \quad (A11)$$

where $T_5 = F_1 F_2 F_3 F_4 F_5$. Then Alice measures her qubits $A_6$ and $A_7$ in the basis $\{|+\rangle, |-\rangle\}$,

and qubit $B$ will be collapsed into the state

$$|\psi_5^+\rangle_B = \frac{1}{2T_5}|\mu^+\rangle_B \quad \text{or} \quad |\psi_5^-\rangle_B = \frac{1}{2T_5}|\mu^-\rangle_B. \tag{A12}$$

If Alice's result of measurement is $|\lambda_4^\perp\rangle_{A_5}$, the qubits $A_6$, $A_7$ and $B$ will be in the state

$$|\phi_5'\rangle = \frac{1}{\sqrt{2}T_5}\left(\frac{y^{16}}{x^{15}}|000\rangle - \frac{x^{16}}{y^{15}}|111\rangle\right)_{A_6 A_7 B}. \tag{A13}$$

Alice can measure her qubit $A_6$ under the basis $\{|\lambda_5\rangle, |\lambda_5^\perp\rangle\}$, which is given by

$$|\lambda_5\rangle = \frac{1}{F_6}\left(\frac{x^{16}}{y^{16}}|0\rangle + \frac{y^{16}}{x^{16}}|1\rangle\right),$$

$$|\lambda_5^\perp\rangle = \frac{1}{F_6}\left(\frac{y^{16}}{x^{16}}|0\rangle - \frac{x^{16}}{y^{16}}|1\rangle\right), \tag{A14}$$

where $F_6 = \left[(x/y)^{32} + (y/x)^{32}\right]^{1/2}$. If Alice's outcome of measurement is $|\lambda_5\rangle_{A_6}$, the qubits $A_7$ and $B$ will be collapsed into the state

$$|\phi_6\rangle = \frac{1}{\sqrt{2}T_6}(x|00\rangle - y|11\rangle)_{A_7 B}, \tag{A15}$$

where $T_6 = F_1 F_2 F_3 F_4 F_5 F_6$. Then Alice measures her qubit $A_7$ under the basis $\{|+\rangle, |-\rangle\}$, and qubit $B$ will be in the state of

$$|\psi_6^+\rangle_B = \frac{1}{\sqrt{2}T_6}|\mu^+\rangle_B \quad \text{or} \quad |\psi_6^-\rangle_B = \frac{1}{\sqrt{2}T_6}|\mu^-\rangle_B. \tag{A16}$$

If Alice's measured result is $|\lambda_5^\perp\rangle_{A_6}$, the state of the qubits $A_7$ and $B$ will evolve as

$$|\phi_6'\rangle = \frac{1}{\sqrt{2}T_6}\left(\frac{y^{32}}{x^{31}}|00\rangle + \frac{x^{32}}{y^{31}}|11\rangle\right)_{A_7 B}, \tag{A17}$$

then she can measure the qubit $A_7$ in the basis

$$|\lambda_6\rangle = \frac{1}{F_7}\left(\frac{x^{32}}{y^{32}}|0\rangle + \frac{y^{32}}{x^{32}}|1\rangle\right),$$

$$|\lambda_6^\perp\rangle = \frac{1}{F_7}\left(\frac{y^{32}}{x^{32}}|0\rangle - \frac{x^{32}}{y^{32}}|1\rangle\right), \tag{A18}$$

where $F_7 = \left[(x/y)^{64} + (y/x)^{64}\right]^{1/2}$. If Alice's outcome of measurement is $|\lambda_6\rangle_{A_7}$, the qubit $B$

will be in the state of

$$\left|\psi_7^+\right\rangle_B = \frac{1}{T_7}\left|\mu^+\right\rangle_B, \tag{A19}$$

where $T_7 = F_1 F_2 F_3 F_4 F_5 F_6 F_7$. If Alice's measured result is $\left|\lambda_6^\perp\right\rangle_{A_7}$, the state of qubit $B$ will evolve as

$$\left|\psi_7^-\right\rangle_B = \frac{1}{\sqrt{2}T_7}\left(\frac{y^{64}}{x^{63}}|0\rangle - \frac{x^{64}}{y^{63}}|1\rangle\right)_B$$

$$= P|\eta\rangle_B, \tag{A20}$$

where $P = \frac{\sqrt{x^{254} + y^{254}}}{\sqrt{2}T_7 x^{63} y^{63}}$, and $|\eta\rangle_B$ is a normalized state, which is given by

$$|\eta\rangle_B = \frac{1}{\sqrt{x^{254} + y^{254}}}\left(y^{127}|0\rangle - x^{127}|1\rangle\right)_B. \tag{A21}$$

Thus, 128 possible final collapsed states of the qubit $B$ are obtained.